\documentclass[aps,pra,superscriptaddress,twocolumn]{revtex4}
\usepackage{dcolumn,graphicx}

\begin{document}

\title{Iterative procedure for computing accessible
  information in quantum communication}

\author{Jaroslav \v{R}eh\'{a}\v{c}ek}
\affiliation{Department of Optics, Palacky University, 17.\ listopadu 50, 
772~00 Olomouc, Czech Republic}

\author{Berthold-Georg Englert}
\affiliation{Department of Physics, National University of Singapore, 
Singapore 117542, Singapore}

\author{Dagomir Kaszlikowski}
\affiliation{Department of Physics, National University of Singapore, 
Singapore 117542, Singapore}

\date{20 August 2004} 

\begin{abstract}
We present an iterative algorithm that finds the optimal measurement for 
extracting the accessible information in any quantum communication scenario.
The maximization is achieved by a steepest-ascent approach toward the extremal
point, following the gradient uphill in sufficiently small steps.
We apply it to a simple ad-hoc example, as well as to a problem with a bearing
on the security of a tomographic protocol for quantum key distribution. 
\end{abstract}


\maketitle

When studying problems in quantum information science, such as investigating
strategies for eavesdropping on quantum communication lines or calculating the
capacity of a quantum channel, one often needs to search for the quantum
measurement that is optimal for the purpose at hand.
Typically, the quantity to be maximized is the accessible information
associated with a set of states.
We describe here a numerical procedure for finding this optimum, and
illustrate it with two examples. 

We consider the following quantum communication scenario.
Alice sends quantum objects to Bob in one of $J$ states,
specified by subnormalized statistical operators $\rho_j$ 
($j=1,\ldots,J$) whose traces
$\tr{\rho_j}$ are the  probabilities which which they are sent.
The state in which Bob receives the objects is therefore given by
\begin{equation}
  \label{eq:A1}
  \rho=\sum_{j=1}^J\rho_j\qquad\textrm{with}
\enskip\tr{\rho}=1\,.
\end{equation}
He performs a positive-operator valued measurement (POVM), composed of $K$
positive operators $\Pi_k$ that decompose the identity,
\begin{equation}
  \label{eq:A2}
  \sum_{k=1}^K\Pi_k=1\qquad\textrm{with}\enskip\Pi_k\geq0\,,
\end{equation}
which he wishes to choose such that the joint probabilities
\begin{equation}
  \label{eq:A3}
  p_{jk}=\tr{\rho_j\Pi_k}\,,\quad\sum_{j,k}p_{jk}=1\,,
\end{equation}
are most informative about the $\rho_j$ that happens to be the case
for the particular quantum object under consideration.
Bob's figure of merit is the \emph{accessible information}
\begin{equation}
  \label{eq:A4}
I=\sum_{j,k} p_{jk}\log\left(\frac{p_{jk}}{p_{j\cdot} p_{\cdot k}}\right)\,,
\end{equation}
where $p_{j\cdot}$ and $p_{\cdot k}$ are the marginal probabilities,
\begin{equation}
  \label{eq:A5}
  p_{j\cdot}=\sum_kp_{jk}=\tr{\rho_j}\,,\quad
  p_{\cdot k}=\sum_jp_{jk}=\tr{\rho\Pi_j}\,.
\end{equation}
Here, $p_{j\cdot}$ is the probability that Alice sends the $j$th state,
$p_{\cdot k}$ is the probability that Bob gets the $k$th measurement outcome, 
and $p_{jk}$ is the probability that she sends the $j$th state \emph{and} he 
gets the $k$th outcome. 

Since Alice's $\rho_j$'s are given, the accessible information of
Eq.~\eqref{A4} is a nonlinear convex functional of Bob's POVM,  
$\Pi\equiv\bigl\{\Pi_1,\ldots,\Pi_K\bigr\}\to I(\Pi)$.
The convexity,
\begin{equation}
  \label{eq:A6}
   I\bigl(\Pi^{(\lambda)}\bigr)\leq
(1-\lambda) I\bigl(\Pi^{(1)}\bigr)+\lambda I\bigl(\Pi^{(2)}\bigr)
\end{equation}
for $\Pi_k^{(\lambda)}=(1-\lambda)\Pi_k^{(1)}+\lambda\Pi_k^{(2)}$ 
with $0\leq\lambda\leq1$, follows from
\begin{equation}
  \label{eq:A7}
  \Bigl(\frac{\partial}{\partial \lambda}\Bigr)^2
  I\bigl(\Pi^{(\lambda)}\bigr)=
\sum_{j,j',k}
\frac{\bigl(p_{jk}^{(1)}p_{j'k}^{(2)}-p_{j'k}^{(1)}p_{jk}^{(2)}\bigr)^2}
{2p_{jk}p_{j'k}p_{\cdot k}}
\geq0\,,
\end{equation}
where $p_{jk}=(1-\lambda)p_{jk}^{(1)}+\lambda p_{jk}^{(2)}$.

Now, since $I(\Pi)$ is convex, it acquires its global maximum at the boundary
of the convex set of all POVMs.
The challenge is then to find the maximizing POVM, and it is the objective 
of this article to describe an iteration procedure for an efficient numerical 
search. 
 
We observe that $I(\Pi)=\sum_k\tr{R_k\Pi_k}$ with
\begin{equation}
  \label{eq:B1}
R_k=\sum_j\rho_j\log\left(\frac{p_{jk}}{p_{j\cdot} p_{\cdot k}}\right)\,,
\end{equation}
and the response of $I(\Pi)$ to a variation of the POVM is given by 
$\delta I=\sum_k\tr{R_k\delta\Pi_k}$ because there is no net contribution 
from the induced changes of the $R_k$'s.
The variations $\delta\Pi_k$ are subject to the constraints of \eqref{A2},
which we enforce by first writing $\Pi_k=A_k^\dagger A_k^{\phdag}$
and then imposing
\begin{equation}
  \label{eq:B2}
  \sum_k\delta\Pi_k=  \sum_k\bigl(A_k^\dagger\delta A_k^{\phdag}
+\delta A_k^\dagger A_k^{\phdag})=0
\end{equation}
on the variations $\delta A_k$.
The most general form for these variations is
\begin{equation}
  \label{eq:B3}
  \delta A_k=i\sum_l \epsilon_{kl}A_l \quad\textrm{with}\enskip
 \epsilon_{kl}^\dagger=\epsilon_{lk}^{\phdag}\,,
\end{equation}
where the $\epsilon_{kl}^{\phdag}$'s are \emph{arbitrary} 
infinitesimal operators.
So,
\begin{equation}
  \label{eq:B4}
  \delta I= i\sum_{k,l}\tr{\epsilon_{kl}^{\phdag}\,A_l^{\phdag}
\bigl(R_k^{\phdag}-R_l^{\phdag}\bigr)A_k^\dagger}\,,
\end{equation}
and the POVMs at the stationary points of $I(\Pi)$ must necessarily be such 
that
$A_l^{\phdag}R_k^{\phdag}A_k^\dagger=A_l^{\phdag}R_l^{\phdag}A_k^\dagger$
or
\begin{equation}
  \label{eq:B5}
  \Pi_l R_k \Pi_k=\Pi_l R_l \Pi_k
\quad\textrm{for all}\enskip k,l\,.
\end{equation}

Upon summing over $k$ or $l$ we arrive at an equivalent set of equations,
\begin{equation}
  \label{eq:C1}
  \Pi_l\Lambda=\Pi_l R_l\,,\quad R_k\Pi_k=\Lambda\Pi_k\,,
\end{equation}
which are adjoint statements of each other because
\begin{equation}
  \label{eq:C2}
  \Lambda=\sum_k R_k\Pi_k=\sum_l \Pi_l R_l
\end{equation}
is hermitian.
Mathematically speaking, $\Lambda$ is the Lagrange multiplier of the constraint
\eqref{B2}, and its physical significance is revealed by noting that
$I=\tr{\Lambda}$. 

The numerical procedure for finding the solution of these equations is an
iteration method that realizes a steepest ascend toward the maximum of
$I(\Pi)$. 
In each round we proceed in the direction of the gradient by putting
$\epsilon_{kl}=-i\alpha A_k^{\phdag}(R_k-R_l)A_l^{\dagger}$ in
\eqref{B3} whereby $\alpha>0$ controls the step size. 
As an immediate consequence of Eq.~\eqref{B4}, the first-order (in $\alpha$)
change of $I(\Pi)$ is then assuredly positive,
\begin{equation}
  \label{eq:C3}
  \left.\frac{\partial I}{\partial\alpha}\right|_{\alpha=0}\!
=\sum_{k,l}\tr{(R_k-R_l)\Pi_k(R_k-R_l)\Pi_l}\geq0,
\end{equation}
and vanishes only at the stationary points where Eq.~\eqref{B5} holds.
But we must correct for the second-order terms that give a nonzero value to
the sum in \eqref{B2}. 

A round of the iteration procedure thus consists of the following three steps. 
\begin{equation}
  \label{eq:iter}
  \parbox[c]{0.8\columnwidth}{
\textbf{Step~1:}~Use the existing approximate POVM to calculate $p_{jk}$, 
$p_{\cdot k}$, and $R_k$ in accordance with Eqs.~\eqref{A3}, \eqref{A5},
and \eqref{B1}.\\
\textbf{Step~2:}~Next, choose a
``small'' positive value for $\alpha$ in 
$G_k=1+\alpha(R_k-\sum_lR_l\Pi_l)$ and compute 
$\tilde{\Pi}_k=G_k^\dagger\Pi_k^{\phdag} G_k^{\phdag}$.\\
\textbf{Step~3:}~Finally, sum up these $\tilde{\Pi}_k$,
 $S=\sum_l\tilde{\Pi}_l$, and  take
$S^{-1/2}\tilde{\Pi}_k S^{-1/2}$
as the new, improved approximation for $\Pi_k$.
}
\end{equation}
We evaluate $I(\Pi)$ at the end of each round to verify that an
acceptable value for $\alpha$ was chosen in Step~2. 
A decrease of $I$, rather than an increase, would indicate an overshooting
and thus tell us that $\alpha$ was too large \cite{speedup}.

The reciprocal square root of $S$ that is needed in Step~3 can be computed
easily and efficiently by a fixpoint iteration, such as the one specified 
by
\begin{equation}
  \label{eq:C4}
  X_{n+1}=\frac{3}{2}X_n-\frac{1}{4}X_n(SX_n+X_nS)X_n\,.
\end{equation}
Starting with $X_0=1$, the sequence of $X_n$ converges rapidly to $S^{-1/2}$, 
provided that all eigenvalues of $S$ are less than $3$.
This condition is surely met if $\alpha$ is small enough because $S-1$ is
positive and proportional to $\alpha^2$.

Any randomly chosen POVM can be used as the starting point for the iteration
\eqref{iter}, except for the maximally ignorant POVMs for which each member
is simply a multiple of the identity, $\Pi_k=p_{\cdot k}$.
Since $p_{jk}=p_{j\cdot} p_{\cdot k}$ for these POVMs, 
they result in $I(\Pi)=0$ and thus mark the global minima of the accessible
information. 
These minima are unstable fixpoints of the iteration \eqref{iter}, and
so it is enough to perturb them slightly by admixing a small fraction of
a randomly generated POVM.

Prior to any iteration, a choice must be made for the value of $K$, that is
the number of elements in the POVM. 
According to a theorem by Davies \cite{Davies:78},
one never needs more than $K=r^2$ members, where $r$ is the rank of $\rho$.
But very often fewer elements will do.
For example, if it is possible to represent all $\rho_j$ by real matrices,
then $K=\frac{1}{2}r(r+1)$ members suffice, as
Sasaki \textit{et al.} have shown \cite{Sasaki+4:99}.
A particular example that has a bearing on the security analysis of certain
schemes for quantum cryptography are \mbox{rank-1} $\rho_j$'s that form a
so-called  ``acute pyramid'', for which one knows that $K=J$ or $K=J+1$ will
do, depending on the volume of the pyramid \cite{pyramids}.

One approach is, therefore, to choose $K=r^2$ for general $\rho_j$'s and 
$K=\frac{1}{2}r(r+1)$ if the $\rho_j$'s have joint real matrix
representations.
After the iterative optimization, one would then look for equivalent members
and combine them into one new member, thereby reducing the value of $K$.
Members $\Pi_{k_1}$ and $\Pi_{k_2}$ are  equivalent if 
$p_{jk_1}p_{j'k_2}=p_{j'k_1}p_{jk_2}$ for all $j$ and $j'$, for then
$R_{k_1}=R_{k_2}$, and the pair $\bigl(\Pi_{k_1}+\Pi_{k_2},0\bigr)$ is as good
as the pair $\bigl(\Pi_{k_1},\Pi_{k_2}\bigr)$. 

Another approach is to begin with a small $K$ value, with $K=J$ suggesting
itself. 
Then, after optimizing for this $K$, one would add a randomly chosen
$\Pi_{K+1}$ and so increase $K$ by $1$, with the proper normalization to unit
sum achieved analogously to Step~3 of \eqref{iter}.
The optimal POVM has been found when the increase of $K$ becomes virtual, 
i.e., when the eventual reduction of equivalent members decreases $K$. 

We note that the iteration scheme \eqref{iter} can be used for the optimization
of other functionals as well, as long as there are given $\rho_j$'s and a
procedure for calculating the $R_k$'s from the $\rho_j$'s and $\Pi_k$'s. 
Indeed, the set of $R_k$'s is the functional gradient of $I(\Pi)$ and can be
regarded as defining $I(\Pi)$ up to a $\Pi$-independent constant.

An example is Helstrom's classic problem of minimum-error discrimination
\cite{Helstrom76},
where one has $K=J$ and wishes to maximize $\sum_jp_{jj}$.
Therefore, we have $R_k=\rho_k$ for Helstrom's problem and can iterate as
described above without further ado.
In view of the very different $R_k$'s, the optimal POVM for minimum-error
discrimination will, as a rule, be different from the POVM that maximizes the
accessible information of Eq.~\eqref{A4}. 
Especially, when the states to be discriminated 
are pure states that are nearly collinear, 
$\tr{\rho_j\rho_{j'}}\lesssim p_{j\cdot}p_{j'\cdot}$,
$\rho$ is almost a pure state itself, and this tends to 
bias the optimal POVM for accessible information 
away from the one that minimizes the discrimination error.
This is well illustrated by the acute pyramids mentioned above \cite{pyramids}.

As a simple example let us consider $J=2$ and $\rho_1$, $\rho_2$ with the
following matrix representations,
\begin{equation}\label{eq:D1}
    \rho_1\widehat{=}\frac{1}{30}\left(
      \begin{array}{ccc}
0 & 0 & 0 \\ 0 & 5 & 2 \\ 0 & 2 & 10
      \end{array}\right)\,,\quad
    \rho_2\widehat{=}\frac{1}{60}\left(
      \begin{array}{ccc}
5 & 2 & 0 \\ 2 & 25 & 0 \\ 0 & 0 & 0
      \end{array}\right)\,.
  \end{equation}
These are rank-2 matrices with $r=3$ for the
rank of $\rho=\rho_1+\rho_2$. 
The Helstrom problem is solved by $\Pi_1$ and $\Pi_2$ projecting on the
subspaces associated with the positive and negative eigenvalues of
$\rho_1-\rho_2$, respectively, 
so that $\max\sum_jp_{jj}=0.840\,888\,4524$. 
As shown in Table~\ref{tbl:expl-Hell} the iteration converges to this value 
in a few dozen rounds.

\begin{table}[t]
\begin{ruledtabular}
\caption{\label{tbl:expl-Hell}%
Hellstrom's Success Rate (SR) $\sum_jp_{jj}$ for the example of
\eqref{D1},
as obtained for a particular iteration sequence that begins with a randomly
chosen $0$th POVM,
given for various Numbers of Iteration Rounds (NIR).
The right column lists the Accessible Information (AI) obtained for the
successive POVMs that optimize SR iteratively.}
\begin{tabular}{rdd}
 \textrm{NIR} &
\multicolumn{1}{c}{\textrm{SR}}
& \multicolumn{1}{c}{AI}\\
\hline
  0    & 0.5                 & 0.0 \\
 10    & 0.838\,445\,2747    & 0.434\,433\,2664 \\
 20    & 0.840\,825\,0481    & 0.447\,716\,5615 \\
 30    & 0.840\,886\,8797    & 0.448\,081\,4579 \\
 40    & 0.840\,888\,4135    & 0.448\,090\,5243 \\
 50    & 0.840\,888\,4515    & 0.448\,090\,7489 \\
$\infty$ & 0.840\,888\,4524 & 0.448\,090\,7546
\end{tabular}
\end{ruledtabular}
\end{table}

\begin{figure}[b]
\centerline{\includegraphics[width=0.9\columnwidth]{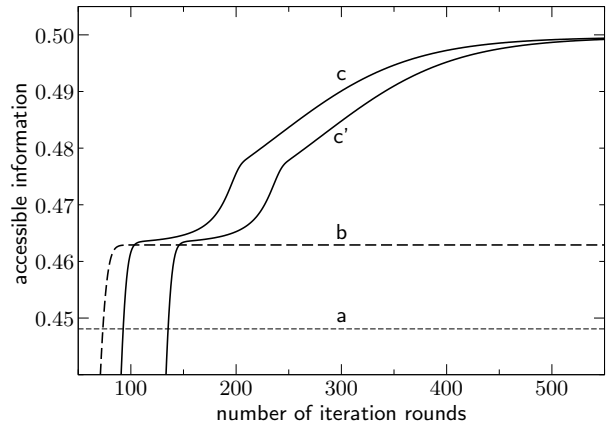}}
\caption{\label{fig:expl-AccInf}%
Values of the accessible information for the example of Eq.~\eqref{D1}, as a
function of the number of iteration rounds. 
Thin dashed horizontal line \textsf{a} shows the value obtained for the POVM 
that optimizes Hellstrom's success rate, see Table~\ref{tbl:expl-Hell}.
Thick dashed line \textsf{b} is for $K=2$, with the POVM of line \textsf{a} 
as the $0$th approximation. 
The optimal POVM is composed of a rank-1 projector and a rank-2 projector.
Curves \textsf{c} and \textsf{c'} are for $K=3$ with different random choices 
for the $0$th approximation.
The optimal POVM is a von Neumann measurement, i.e., it is composed of three
rank-1 projectors.
}
\end{figure}

The respective iterations for maximizing the accessible information are
summarized in Fig.~\ref{fig:expl-AccInf} for $K=2$ and $K=3$.  
No improvement is found for $K=4$, $5$, or $6$, so that the optimal POVM has
three members in this example.
No attempt has been made, for Table~\ref{tbl:expl-Hell} or
Fig.~\ref{fig:expl-AccInf}, to optimize parameter $\alpha$ of \eqref{iter}.
A judicious choice could reduce the necessary number of iteration rounds
by much. 

The algorithm of \eqref{iter} provides \emph{numerical} answers 
for the accessible information and identifies the optimal POVMs.
This is already sufficient for many quantum information tasks,
such as the checking of the security of noisy quantum communication,
or the calculation of the maximal information yield per quantum system sent.
The applications are not limited to these, however.
Sometimes it is possible, upon restricting the set of measurements
over which the accessible information is maximized to some special
class of measurements, to derive analytical forms for the optimal
POVM. 
But its \emph{global} optimality can often be only conjectured, 
without having a solid proof. 
Our algorithm can be used, in such cases, 
to check the optimality of the POVM found.
Such checks were indeed performed for the low-dimensional ``pyramids'' of
Ref.~\cite{pyramids}. 

Another, even more important potential application is
the  use of the algorithm for finding the \emph{analytical} 
solutions to optimization problems. 
For instance, it is typical in quantum communication protocols
that the states that are to be distinguished by the eavesdropper 
constitute a highly symmetric family of quantum states.
The optimal POVM then also tends to be relatively simple, which makes it 
sometimes possible to reconstruct its analytical form from the result of 
the numerical search. 

As an illustration, we consider the qubit version of the fully tomographic
protocols for quantum key distribution of Ref.~\cite{tomocrypt}, an extension
of the classic 4-state protocol of Bennett and Brassard \cite{BB84}, or rather
its 6-state generalization \cite{6state}, with some features of Ekert's
protocol \cite{E91}, which itself involves partial tomography.
In this protocol, eavesdropper Eve controls the source that distributes qubit
pairs to the Alice and Bob, one pair at a time, in such a way that the
statistical operator of these pairs is the singlet state $\ket{\psi}$
with an admixture of unbiased noise,
\begin{equation}
  \label{eq:E1}
  \rho_\mathrm{A\&B}=\ket{\psi}(1-\epsilon)\bra{\psi}+\frac{\epsilon}{4}
\end{equation}
with $0\leq\epsilon\leq1$.

Both Alice and Bob measure the six-element POVM consisting of the rank-1
operators $\frac{1}{6}(1\pm\sigma_\zeta)$ with $\zeta=x,y,z$, where the
$\sigma_\zeta$'s are the respective basic Pauli operators for their qubits. 
The mutual information of the resulting joint probabilities is
\begin{equation}\label{eq:E2}
I_\mathrm{A\&B}=\frac{1}{6}\Bigl[\epsilon\log_2 \epsilon+
(2-\epsilon)\log_2(2-\epsilon)\Bigr]\,,
\end{equation} 
where we now adopt the conventions of information theory and employ the binary
logarithm rather than the natural logarithm that is more convenient in
Eqs.~\eqref{A4}---\eqref{iter}. 

At the source, each pair is entangled with an ancilla, which remains in 
Eve's possession as a quantum record of what has been sent to Alice and Bob.
As shown in Ref.~\cite{tomocrypt} (see also the appendix in \cite{pyramids}),
the best Eve can do is to use another qubit pair for the ancilla and prepare
the pure state
\begin{equation}
  \label{eq:E3}
  \ket{\Psi}=\ket{\psi_{12}\psi_{34}}a+\ket{\psi_{13}\psi_{24}}b\,,
\end{equation}
where qubits $1$ and $2$ are Alice's and Bob's, respectively, qubits $3$ and
$4$ make up the ancilla, and $\psi_{jk}$ means the singlet for qubits $j$ and
$k$.
The constraint $\tr[34]{\ketbra{\Psi}}=\rho_\mathrm{A\&B}$ requires
\begin{equation}
  \label{eq:E4}
  \bigl|2a+b\bigr|^2=4-3\epsilon\,,\quad
  \bigl|b\bigr|^2=\epsilon\,,
\end{equation}
and it is convenient to choose the arbitrary complex phases of $a$ and $b$
such that ${\bar{\epsilon}\equiv(2a+b)b^*=\sqrt{4\epsilon-3\epsilon^2}}$ 
is positive.

For each of Alice's (or Bob's) six measurement results there is a corresponding
ancilla state,
\begin{equation}
  \label{eq:F1}
  \rho_{\zeta\pm}
=\frac{1}{24}\biggl[1\mp\frac{1}{2}(\epsilon+\bar{\epsilon})\sigma^{(3)}_\zeta
                    \mp\frac{1}{2}(\epsilon-\bar{\epsilon})\sigma^{(4)}_\zeta
              -(1-\epsilon)\vec{\sigma}^{(3)}\cdot\vec{\sigma}^{(4)}\biggr]
\end{equation}
with $\zeta=x,y,z$.
For Eve it is, therefore, as if Alice were sending her these subnormalized
statistical operators, and what Eve can know about Alice's measurement results
is measured by the accessible information associated with this sextet. 

Eve's optimal POVM can be found numerically by the iteration procedure of
\eqref{iter}.
One finds that it has six elements which, upon careful inspection of the
numerical results, are identified as the rank-1 operators
\begin{equation}
  \label{eq:F2}
  \Pi_{\zeta\pm}=\frac{1}{6}
\biggl[1\mp\frac{\sqrt{3}}{2}\bigl(\sigma^{(3)}_\zeta-\sigma^{(4)}_\zeta\bigr)
-\frac{3}{2}\sigma^{(3)}_\zeta\sigma^{(4)}_\zeta
+\frac{1}{2}\vec{\sigma}^{(3)}\cdot\vec{\sigma}^{(4)}\biggr]
\end{equation}
with $\zeta=x,y,z$.
Note the remarkable simplicity of the optimal POVM: it has relatively few
elements, and does not depend on the noise parameter $\epsilon$. 
This is clearly a consequence of the symmetry of the sextet \eqref{F1}, which
consists of unitarily equivalent rank-2 operators.

The accessible information gained by Eve from this POVM is given by
\begin{equation}
  \label{eq:F3}
  I_\mathrm{A\&E}
 =I_\mathrm{A\&B}\bigl(\epsilon\to 1-\sqrt{3/4}\,\bar{\epsilon}\bigr)
\end{equation}
with $I_\mathrm{A\&B}(\epsilon)$ as in Eq.~\eqref{E2}.
Accordingly, the critical $\epsilon$ value, 
for which $I_\mathrm{A\&E}=I_\mathrm{A\&B}$,
is given by
\begin{equation}
  \label{eq:F4}
  \epsilon_\mathrm{crit}=\Bigl(\frac{5}{2}+\sqrt{3}\Bigr)^{-1}
\simeq 0.2363\,.
\end{equation}
For $\epsilon<\epsilon_\mathrm{crit}$, 
we have $I_\mathrm{A\&B}>I_\mathrm{A\&E}$, so that
the information that Bob has about Alice's bit values exceeds Eve's
information about them. 
Then, the  Csisz\'ar--K\"orner theorem \cite{CK} ensures that they can 
generate a private, secure cryptographic key by one-way communication.   

We must not fail to mention that the POVM of Eq.~\eqref{F2} is only optimal 
for $\epsilon<\frac{2}{3}$, that is in the interesting parameter range where 
$\rho_\mathrm{A\&B}$ is not separable and the joint probabilities between
Alice and Bob contain nonclassical correlations.
For $\epsilon\geq\frac{2}{3}$, Eve can blend $\rho_\mathrm{A\&B}$ from product
states and thereby obtain $I_\mathrm{A\&E}=\frac{1}{3}$ right away. 

In conclusion, we have presented an iterative numerical procedure for finding
the POVM that optimally extracts the accessible information from a given set
of states received in a quantum communication scenario. 
The method is a steepest-ascend approach toward the maximum; it follows the
gradient in steps that are suffiently small to avoid over-shooting, so that
the accessible information increases monotonically in each iteration step.
We have illustrated the method at a simple ad-hoc example.
A second example, which has a bearing on quantum key distribution,
shows how the analytical answer can be established,
once crucial insight is gained from the numerical solution.
  
\vfill

We wish to thank Frederick Willeboordse for valuable discussions. 
This work was supported by 
Grant No.~LN00A015 of the Czech Ministry of Education,
by A$^*$Star Grant No.\ 012-104-0040,
and by NUS Grant WBS: R-144-000-089-112.


\begin{thebibliography}{99}
\bibitem{speedup}
One may speed up the convergence by adjusting the value of $\alpha$.
For example, after finding the new values $I_1$, $I_2$ of $I(\Pi)$ for two
trial values $\alpha_1$, $\alpha_2$, a good choice is 
\\[0.1\baselineskip] \centerline{$\displaystyle
\alpha=\frac{1}{2}\frac{(I_1-I_0)\alpha_2^2+(I_0-I_2)\alpha_1^2}
{(I_1-I_0)\alpha_2+(I_0-I_2)\alpha_1}$}\\[0.1\baselineskip]
where $I_0$ is the old value of $I(\Pi)$.
\end{thebibliography}
\end{document}